\documentclass[aps,prd,showpacs,preprintnumbers]{revtex4}
\usepackage{amsmath,amssymb}
\usepackage{slashed}
\usepackage{graphicx}
\usepackage{subfigure}
\usepackage{color}

\newcommand\bef{\begin{figure}}
\newcommand\eef[1]{\label{fig.#1}\end{figure}}
\newcommand\beq{\begin{equation}}
\newcommand\eeq[1]{\label{eq.#1}\end{equation}}
\newcommand\beqa{\begin{eqnarray}}
\newcommand\eeqa[1]{\label{eq.#1}\end{eqnarray}}
\newcommand\bet{\begin{table}}
\newcommand\eet[1]{\label{tbl.#1}\end{table}}
\newcommand\betb{\begin{center}\begin{tabular}}
\newcommand\eetb{\end{tabular}\end{center}}
\newcommand\beit{\begin{itemize}}
\newcommand\eeit{\end{itemize}}

\newcommand\fig[1]{Fig. \ref{fig.#1}}
\newcommand\eq[1]{Eq.\ (\ref{eq.#1})}

\newcommand{\qq}{\bar{Q}Q}

\newcommand{\jpsi}{J/\psi}

\newcommand{\yna}{\Upsilon(1S)}

\newcommand{\md}{m_{\scriptscriptstyle D}}
\newcommand{\lqcd}{\Lambda_{\scriptscriptstyle QCD}}
\newcommand{\eb}{\epsilon_{\scriptstyle B}}

\newcommand{\nc}{N_c}

\newcommand{\lnr}{\mathcal{L}_{\scriptscriptstyle NR}}
\newcommand{\lqq}{\mathcal{L}_{\scriptscriptstyle Q}}
\newcommand{\lpnr}{\mathcal{L}_{\scriptscriptstyle pNR}}

\newcommand\tr{\mathrm{Tr}\;}

\newcommand\epj{{\sl Eur.\ Phys.\ J.\/}\ }
\newcommand\jhep{{\sl J.\ H.\ E.\ P.\/}\ }
\newcommand\np{{\sl Nucl.\ Phys.\/}\ }
\newcommand\npbps{{\sl Nucl.\ Phys.\/} B ({\sl Proc. \ Suppl.\/})\ }

\newcommand\pr{{\sl Phys.\ Rev.\/}\ }
\newcommand\pls{{\sl Phys.\ Lett.\/}\ }

\newcommand\ijmp{{\sl Int.\ J.\ Mod.\ Phys.\/}\ }
\begin{document}
\title{On decay width of heavy quarkonia in QGP}
\author{Saumen Datta}
\email{saumen@theory.tifr.res.in}
\affiliation{Department of Theoretical Physics, Tata Institute of Fundamental
         Research,\\ Homi Bhabha Road, Mumbai 400005, India.}
\begin{abstract}
Quarkonia are some of the most important probes of the medium created in
relativistic heavy ion collision experiments, but it is still difficult to get
quantitative results for its behavior in the plasma. Here I discuss the
decay width of a heavy $\qq$ system, and calculate the gluodissociation 
width of bottomonia. In the end I comment on study of quarkonia as 
open quantum systems.
\end{abstract}

\maketitle

\section{Introduction}
\label{sec.intro}
Quarkonia, mesons of heavy quark and antiquark, constitute one of the
most popular probes of quark-gluon plasma. In the plasma the reduced binding 
between quark and antiquark and the presence of energetic thermal
gluons lead to a reduction in yield of a $Q \bar{Q}$ meson
\cite{matsui}. Since the vector quarkonia like $\jpsi$ and $\yna$ 
can be clearly detected through their decay into dileptons, it has
become one of the most studied observables in relativistic heavy
ion collisions, both experimentally and theoretically 
\cite{review}.

While the large mass of the quark provides various simplifications in
the theory side, quantitative predictions have remained difficult, at
least in the temperature range of interest for heavy ion collision
experiments. Direct lattice studies are difficult due to the
requirement of analytic continuation to connect to experimental
observables. Early lattice results \cite{asakawa,datta} predicting
very small temperature effects on $\jpsi$ yield has been questioned 
\cite{mocsy}, and a later study has suggested substantial thermal
modification on crossing $T_c$ \cite{ding}. For bottomonia,
nonrelativistic effective field theory on lattice has been employed;
however, there is a wide spread in the thermal width estimates
\cite{aarts,kim}. On the other hand, very interesting theoretical
insights have been obtained, from use of various effective field
theory techniques \cite{laine,brambilla}, as well as from using 
concepts of open quantum systems \cite{akamatsu,blaizot}. However, 
most of thse works are based on perturbation theory, and therefore, it
is difficult to extract quantitatively accurate predictions from them.

In this note I discuss use of lattice studies to complement the
effective field theory works, and in particular, use it to calculate the 
decay width of $\yna$ in plasma. Then I make some general observations 
about constraining open quantum system studies.  

\section{Heavy quarkonia in QGP}
\label{sec.decay}

Since we will work with heavy quarks whose mass is larger than all
other scales in the theory including the temperature scale, it is
natural to start with the nonrelativistic action,
\beqa
\lnr &=& \lqq \ + \ \bar{\psi} 
\left( i \slashed{D} - m \right) \psi \ - \ \frac{1}{2} \ \tr G_{\mu \nu}  
\, G_{\mu \nu}  \nonumber \\
\lqq &=& \phi^\dag \left( i D_0 + 
\frac{{\bf D}^2}{2 M} \right) \phi \ + \ \chi^\dag \left( i D_0 
- \frac{{\bf D}^2}{2 M} \right) \chi + ...
\eeqa{NR}
where ${\phi,\chi}$ are the two-component fields that annihilate heavy 
quark and antiquark, respectively, and $\lqq$ is the nonrelativistic 
heavy quark part of the action. The other parts in \eq{NR} relate to
the light quark action and the gauge action, respectively, and $\lnr$
is an effective lagrangian for energy scales $\ll M$. 
We will use the generic term ``heavy 
quarks'' to denote a configuration of heavy quarks and antiquarks.

If we further assume that temperature scale is much larger than all
other scales in the theory, $T \gg M v, \lqcd$, then one can integrate
out this scale and get a complex potential to describe the $\qq$ pair
\cite{laine}. The real part of the potential is the 
well-known debye-screened Yukawa potential, and the imaginary part 
describes decay of the $\qq$ meson via Landau damping
\cite{blaizot2,laine2}. While the real part of the potential can be
extracted from nonperturbative lattice studies with reasonable control 
\cite{burnier}, the imaginary part is much more difficult to obtain. 
Note also that the scale ordering $T \gg Mv$ is not valid in the
context of bottomonia in relativistic heavy ion collision. A
systematic study of different scale orderings was carried out in
Ref. \cite{brambilla}, where it was shown that for the physically more
interesting case $Mv \gtrsim T$ the leading mechanism of the decay of the
$\qq$ meson is gluodissociation.

The main purpose of this note is to make a nonperturbative estimate of the
gluodissociation width. We will do that in the leading order of $\lqq$
in \eq{NR}, and assuming the scale hierarchy 
\beq
M v \sim 1/r \gg T \gtrsim \md \gg \eb,
\eeq{scale} 
where $\eb$ is the binding energy of the $\qq$ meson.

The interaction of static $\qq$ singlet with a gluonic field has
already bin worked out by Peskin \cite{peskin}, who showed that
the interaction of the gluonic field with the $\qq$ pair is like a
color electric dipole term: summing over the time evolution, one gets \\
\beq
\frac{g^2}{2 \nc} \int dt \int_0^\infty d\tau \ 
\langle \vec{r}.\vec{E_a}(t) \ e^{-(H_o - H_s) \tau}
\ \vec{r}.\vec{E_a}(t-\tau) \rangle
\eeq{gluod}
where $H_o, H_s$ are the hamiltonians for the octet and singlet,
respectively, and $E_a$ is the color electric field. 

To find the contribution of this term to the decay width of
$\yna$, following Peskin we write the $\yna$ state formally as 
$\vert \Upsilon(R) \rangle \equiv \vert R \rangle \ \vert \mathbb{I}_c
\rangle \ \vert \psi(r) \rangle$, where $R$ denotes the c.m. and
$\psi(r)$ corresponds to the wavefunction in relative
coordinates. Also (following Peskin) we can set the energy of the
adjoint state to $0$ (compared to free particle state) and so the
energy difference in the exponential $\sim \eb$.

In the vacuum cross-section calculation, the energy exponential plays a
crucial role in the total matrix element. On the other hand, here $E
E$ thermal correlator is expected to have a range $\sim 1/\md$, and
$\md \gg \eb$ (\eq{scale}). So we can ignore the effect of this term
(it is of the same order as subleading terms) \cite{brambilla2}. Then we get 
\beqa
\Gamma_g \ &=& \ 2 \frac{g^2}{2 \nc} \langle \phi \vert r_i r_j \vert \phi
\rangle \ \int d\tau \langle E^a_i(\tau) E^a_j(\tau) \rangle_T
\nonumber \\
&=& \frac{g^2}{6 \nc} \int d\tau \langle E^a_i(\tau) E^a_j(\tau) \rangle_T
\ \int d^3r \, \phi(r)^2 r^2 
\eeqa{gamma}
where $\phi(r)$ is the spatial wavefunction of the $\yna$.    

The thermal matrix element in \eq{gamma} has already been calculated
on lattice in Refs. \cite{francis,banerjee} in the context of
study of momentum diffusion coefficient of heavy quarks in plasma. 
The momentum diffusion coefficient, $\kappa$, is defined through a 
Langevin equation for a heavy quark in plasma \cite{svetitsky,moore}. 
\beqa
\frac{d \vec{p}}{d t} &=& - \frac{1}{2 MT} \, \gamma \, \vec{p} \ + \ \xi(t) 
\nonumber \\
\langle \xi_l(t) \, \xi_m(t^\prime) \rangle \ &=& \ \kappa \; \delta_{lm} \; 
\delta(t-t^\prime).
\eeqa{langevin}
In field theory, the corresponding quantity of interest is 
$M \dot{\vec{J}}$, where $\vec{J}$ is the number density current. 
Using \eq{NR} one gets
\beq
M \dot{\vec{J}} = \phi^\dagger \; g \vec{E} \; \phi \ - \ \chi^\dagger \; 
g \vec{E} \; \chi.
\eeq{current} 
Using fluctuation-dissipation theorem and some manipulation one can then 
show that \cite{caron-huot} 
\beq
\kappa \ \propto \ \lim_{\omega \to 0} \ \int dt \; e^{i \omega t} \int dx \ \left\langle \left\{ 
M \dot{\vec{J}}(t,\vec{x}),\ M \dot{\vec{J}}(0,0) \right\} \right\rangle.
\eeq{kappa}
On the lattice, one can calculate the Matsubara correlator of the 
electric field, and extract $\kappa$ \cite{caron-huot,francis,banerjee}: \\
\beq
\kappa \ = \ \lim_{\omega \to 0} \ \frac{2 T}{\omega} \ \rho(\omega) \cdot
\eeq{lattice}
Here $\rho$ is the spectral function for the electric field operator.

To calculate the spatial matrix element, we solve for
the ground state singlet wave function $\phi(r)$ using the real part
of the $\qq$ potential. As mentioned before, this potential has been
calculated on the lattice \cite{burnier}. We use the parametrization
of the potential given in \cite{chen} \footnote{Ref. \cite{chen} uses the
  form of the 1-dimensional screened cornell potential, 
\[
V(r,T)= - \frac{\alpha}{r} e^{- \md r} \ + \ \frac{\sigma}{\md}
\ \left( 1 - \exp(-\md r) \right)
\]
where $\md$, the Debye mass, is fitted to the form 
\[
\frac{\md}{T} = a (T/T_c-b)^c + d.
\] 
Using the lattice results of \cite{burnier}, and setting $T_c$ = 172.5
MeV, they obtain a=6.32, b=0.885, c=0.1035 and d=-4.058.} and take 
as vacuum potential the Cornell form $V(r)=- \frac{\textstyle
  \alpha}{\textstyle r} \, + \, \sigma r$, with $\alpha$=0.3872,
$\sigma$ = 0.2025 GeV$^2$, and $m_b$ = 4.68 GeV.

In \fig{kappagamma} I show our estimates for the gluodissociation
width thus obtained. In the left hand figure is shown the calculation
of the momentum diffusion coefficient $\kappa$ in SU(3) gluon plasma. 
This is a slightly updated figure from that in Ref. \cite{banerjee}: for the 
renormalization constant the one-loop result of \cite{christensen} is used.
Also here we only show the statistical (including fitting) errorbar.  
Here we are using this calculation only to get the decay width, which has its 
own, large but different, set of systematic errors; at the moment we are not 
making an attempt to make a serious assesment of the systematic error. 
One such obvious issue is the fact that the lattice measurement of $E-E$ 
correlator was done for a gluon plasma, with $T_c \sim$ 260 MeV. We note,
however, that plotted in units of $T_c$, $\kappa$ agrees well with 
experimental measurements of the quantity. Encouraged by this, for our 
estimate we assume that $\kappa/T_c^3$ has similar values for full QCD at 
similar value of $T/T_c$, and expect the estimate of $\gamma/T_c$ to 
be reasonably good also for QCD. For estimate of $\Gamma$ in MeV, one can use 
$T_c$ =172.5 MeV.

\bef
\centerline{\includegraphics[scale=0.7]{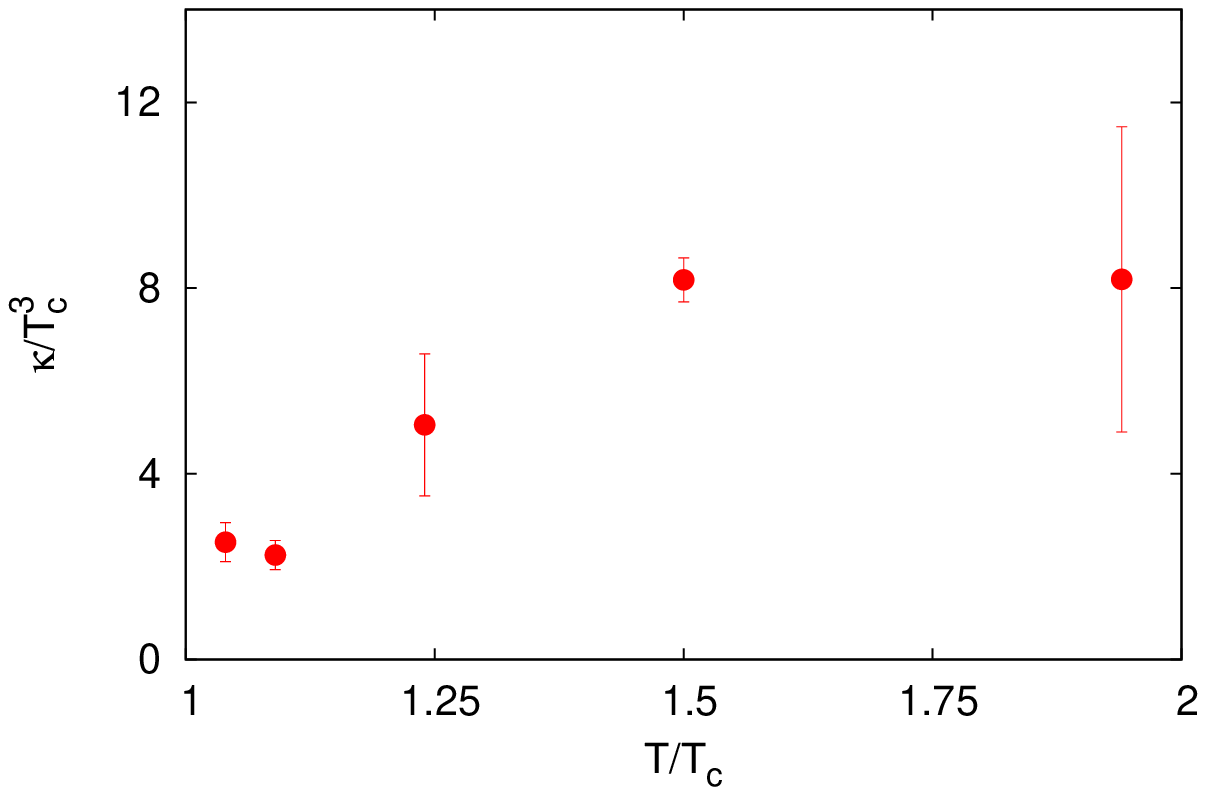}\includegraphics[scale=0.7]{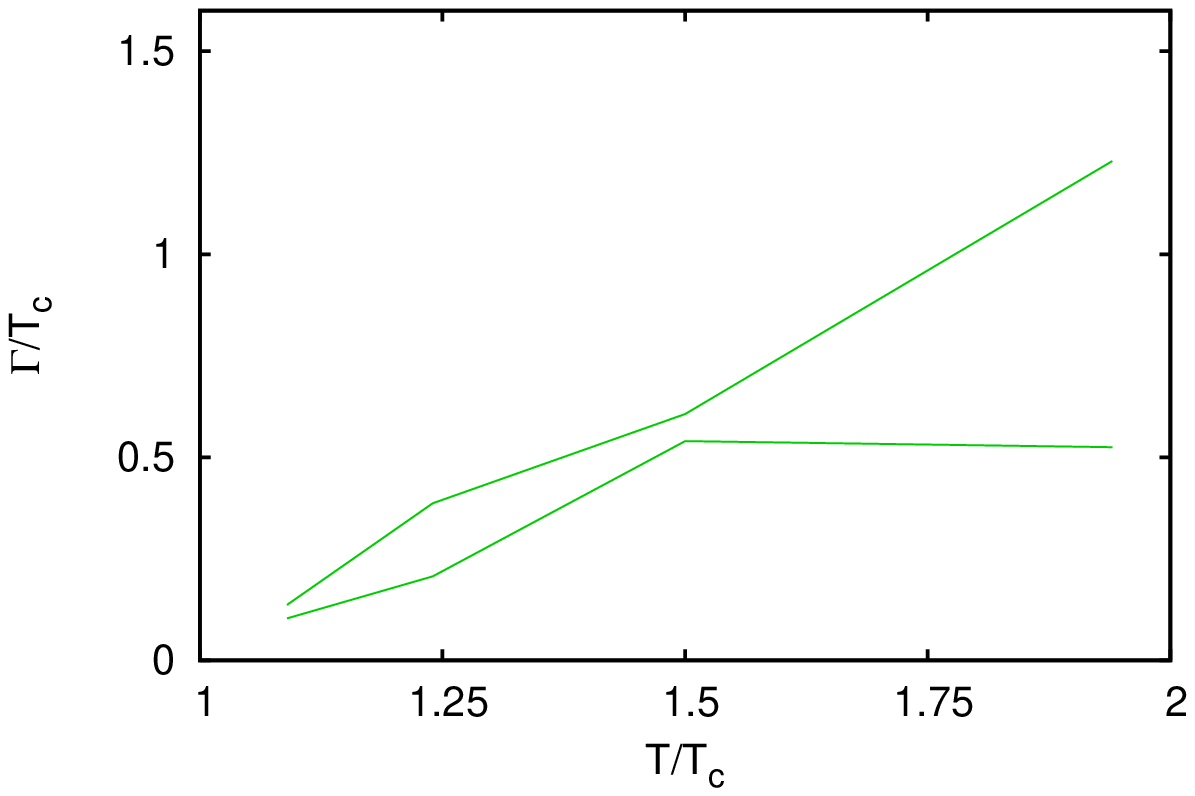}}
\caption{(left) Momentum diffusion coefficient $\kappa$ in SU(3) gluon plasma
(data from \cite{banerjee}). The error bar shown is only statistical. 
(right) The decay width for the state $\yna$, shown in unit of $T_c$. See text.
The area covered by the two lines show estimate of the 1-$\sigma$ statistical 
band.}
\eef{kappagamma}
      
Before we discuss our result, a few comments are in order. The connection of 
decay width of $\qq$ system to the electric field correlator, at a similar 
level of approximation,  was discussed 
first in Ref. \cite{brambilla2}. Using the pNRQCD effective field theory 
approach, they came directly to the electric field electric field correlator
(see also \cite{brambilla}). In pNRQCD one writes an effective 
lagrangian for the $\qq$ system. The pNRQCD lagrangian is

\beq
\lpnr = \int d^3r \ {\rm Tr} \left\{ S^\dagger \left( i \partial_0 - V_s \right) S
+ O^\dagger \left( i D_0 - V_o \right) O \right\} + Z_A(r) \ \tr \left\{ O^\dagger
\; \vec{r}.g\vec{E} \; S \ + \ S^\dagger   \; \vec{r}.g\vec{E} \; 
O \right\} + ...
\eeq{pNR}

where $S,O$ refer to octet and singlet configurations of $\qq$, $r$ is the 
relative coordinate, and $Z_A$, the matching coefficient, is 1 in leading order
(see \cite{pnr} for a review). The dipole interaction vetices $\vec{r}.\vec{E}$
connect the singlet to the octet. Therefore one immediately gets \eq{gluod}. 
The difference between their work and ours is in our treatment of the spatial
part $\langle \phi \vert r^2 \vert \phi \rangle$, which will lead to a 
different behavior with temperature since this factor changes quite 
substantially in the temperature range we have discussed.

Another recent estimation of gluodissociation width was made in 
\cite{chen}. While our treatment of the spatial wavefunction is 
similar to theirs, they used the perturbative estimate of \cite{peskin} for 
the electric field correlator, leading to an order-of-magnitude smaller 
width at comparable temperatures. Their approach leads to the somewhat 
counterintuitive result that the decay width starts decreasing with 
temperature after $\sim 1.4 T_c$. 

Now let us look at the results obtained in \fig{kappagamma}. The decay width 
of $\yna$ is small at small temperatures, but rises quite fast with 
temperature, reaching $\sim$ 100 MeV by 1.5 $T_c$. Note that this is quite 
a large width, since the plasma lasts for almost 10 fm. However, it is 
considerably smaller than the estimate made in \cite{aarts} from direct 
lattice studies. Due to the large 
statistical error it is difficult to identify a trend at higher temperatures,  
but over this temperature range a linear rise with $T-T_c$ is consistent with 
data within error. Of course, this linearity is the result of combined effect 
of different factors; at very high temperatures one expects a $\sim T^3$ 
behavior with temperature \cite{brambilla}. It will be interesting to see 
if the approach to such behavior sets in at moderately high temperatures.

\section{Heavy quark system in plasma}
\label{sec.HQ}

In the previous section we connected the calculation of decay width of 
$\yna$ to the motion of a meandering $b$ quark. This is a pointer to the idea 
that rather than thinking about the quarkonia separately, it may be more 
useful to think of the heavy quark system as a whole, and its interactions 
with the plasma. More generally, treating the problem of quarkonia in plasma 
as an open quantum system has become popular, see, e.g., 
\cite{akamatsu,blaizot}. Setting of the pNRQCD effective theory in the
open quantum system setup has also been considered in \cite{brambilla2}.
Phenomenological study of quarkonia within the open quantum system framework 
has also been considered \cite{kajimoto,rishi}.  
 
We do not intend to go to the machinery of open quantum system here. 
We will, however, see how Section \ref{sec.decay} sits within a more 
general framework introduced in \cite{blaizot}. 
Starting from a configuration of heavy quarks ${Q_i}$ at time $t_i$, the 
probablitiy of finding the heavy quark system in a configuration
${Q_f}$ at time $t_f$ can be written as \cite{akamatsu,blaizot}

\beq 
P\left[{Q_f},t_f \vert {Q_i}, t_i \right] \ = \ \int \mathcal{DQ} \, \int
\mathcal{D}[\bar{\psi},\psi] \ \exp(i S[Q, \psi]) \ \equiv \ \int 
\mathcal{D}[\phi, \chi] \ e^{i \Phi[A]}
\eeq{prob}

where the influence functional, $\Phi[Q]$ \cite{akamatsu,blaizot}
\beq
e^{i \Phi} \ = \ \int \mathcal{D}A_0 \ e^{-i \int \rho A_0} \ e^{i S^\prime}
\eeq{Phi}
includes all the terms with light quarks and gluons.
One can write a path integral expression for $\Phi$ using the
well-known Schwinder-Keldysh contour.

It can be shown \cite{blaizot} that the above path integral can be 
obtained from the generalized Langevin equation \\
\beq
M \frac{d^2}{dt^2} {\bf R} \ = \ - M \gamma(R) \dot{\bf R} \ - \nabla_R
V(R) \ + \ \xi(R, t)
\eeq{langevinm}
by averaging over $\xi$. $\xi$ is a white noise: 
\beq
\langle \xi_l(R,t) \, \xi_m(R, t^\prime) \rangle \ = \ \kappa_{lm}(R)
\; \delta(t-t^\prime). 
\eeq{white}
The 2n-component column vector $R$ has the position vectors of
the n quarks and antiquarks, 
${\bf R}^T = \left(\vec{r_i}, \vec{\bar{r_i}} \right)$ and 
$V(R)$ is the potential due to other heavy quarks and antiquarks.
$\nabla_R V(R)$ is a shorthand for interaction terms between all possible pairs:
it is a column vector with terms like 
\[ \nabla_R V(R) = \begin{bmatrix} \nabla V(r_i-r_j)-\nabla V(r_i-\bar{r}_j), \\ 
                                   ..., i=1,n, \\
                                  \nabla V(\bar{r}_i-\bar{r}_j)-\nabla V(\bar{r}_i-r_j), \\
..., i=1,n
\end{bmatrix} 
\]
while the matrix $\gamma$ is $2n \times 2n$ matrix with elements like 
$H(r_i-r_j)=\nabla_{kl} W(r_i-r_j)$, with $W$ the imaginary part of the 
2-body potential. 
The medium information is included in the noise term. Note that
an individual collision with the medium particles with momenta $\sim T$ imparts
a momenta kick $\lesssim T$, which does not change the heavy particle
momentum $P \sim \sqrt{2 M T}$ substantially, therefore justifying the 
white noise assumption \cite{svetitsky,moore}.

If we take an isolated heavy quark, sufficiently far away from all the 
other quarks and antiquarks, then for it the $V(R)$ term does not contribute 
and we get back \eq{langevin}.
On the other hand for $Q \bar{Q}$ pair at separation
$r \ll T$,  from \eq{langevinm} one
gets the equation of the pair 
under the influence of a mutual potential and scattering from the 
electric gluons.

This generic formalism generalizes the connection of Section \ref{sec.decay}. 
Similarly, this also opens up the possibility of parametrizing the theory of the
generic open quantum system using studies of isolated quarkonia on the lattice.

\end{document}